\documentclass[12pt]{article}
\usepackage{amsmath, amsfonts,euscript,bbm}
\usepackage{graphicx}

\def\be{\begin{equation}}
\def\ee{\end{equation}}
\def\beq{\begin{eqnarray}}
\def\eeq{\end{eqnarray}}
\def\bea{\begin{eqnarray}}
\def\eea{\end{eqnarray}}

\def\gev{\, {\rm GeV}}

\newcommand{\gsim}{\lower.7ex\hbox{$\;\stackrel{\textstyle>}{\sim}\;$}}
\newcommand{\lsim}{\lower.7ex\hbox{$\;\stackrel{\textstyle<}{\sim}\;$}}

\begin{document}

\setlength{\baselineskip}{0.25in}

\begin{titlepage}
\noindent
\begin{flushright}
MIFP-07-09 \\
hep-th/0703278
\end{flushright}
\vspace{1cm}

\begin{center}
  \begin{Large}
    \begin{bf}
An Inflationary Scenario in Intersecting Brane Models \\

    \end{bf}
  \end{Large}
\end{center}
\vspace{0.2cm}
\begin{center}
\begin{large}
Bhaskar Dutta, Jason Kumar and Louis Leblond \\
\end{large}
  \vspace{0.3cm}
  \begin{it}
Department of Physics, Texas A\&M University \\
        ~~College Station, TX  77843, USA \\
\vspace{0.1cm}
\end{it}
\end{center}

\begin{abstract}
We propose a new scenario for $D$-term inflation which appears
quite straightforwardly in the open string sector of intersecting brane
models.  We take the inflaton to be a chiral field in a bifundamental
representation of the hidden sector and we argue that a sufficiently flat potential can be
brane engineered. This type of model generically predicts
a near gaussian red spectrum with negligible tensor modes. We
note that this model can very naturally generate a baryon
asymmetry at the end of inflation via the recently proposed
hidden sector baryogenesis mechanism.  We also discuss the
possibility that Majorana masses for the neutrinos can be
simultaneously generated by the tachyon condensation which
ends inflation. Our proposed scenario is viable for both high and low scale supersymmetry
breaking.
\end{abstract}

\vspace{1cm}

\begin{flushleft}
March 2007
\end{flushleft}

\end{titlepage}

\setcounter{footnote}{0}
\setcounter{page}{2}
\setcounter{figure}{0}
\setcounter{table}{0}


\section{Introduction}

For many years, one of the focusses of string theory has
been an attempt to find realistic compactifications which
could potentially make contact with low-energy phenomenology.
Much of the original work on this subject dealt with
compactifications of heterotic string theory.  But the more
recent understanding of non-perturbative string theory and
the role of D-branes has allowed a shift of focus to
Type IIA/B string compactifications.  In particular, one of
the current promising avenues for studying Standard Model
embeddings in string compactifications are Type II intersecting brane
models \cite{IBM,IBMFV}.

More recently, beginning with the surprising observation of a positive
cosmological constant, there have been increasing attempts to
find string models which can match current cosmological observations
and theories \cite{reviewinflation}.  Particular attention
has been paid to the question of
how inflation can be realized in string constructions.
Most of the recently constructed models have been shown
to be fine-tuned in one way or another through
the well known supergravity $\eta$ problem.
Until now, there has been relatively little effort aimed at bringing
together these two strains by constructing string models which admit
Standard Model physics at low-energy, with potentials which can
exhibit inflation in the early universe.
One of the most common attempts to embed
inflation in string theory arises in brane inflation \cite{stringinf, BBbar, dbi}.
In the most recent realistic implementation of these models \cite{kklmmt} (see \cite{Bean:2007hc}
for the most recent comparison to data),
the inflaton is an open string scalar which describes the position of
a (anti)D3-brane which is moving down a warped throat. The Standard Model is
assumed to arise from open-string dynamics involving D-branes in a different
throat (see \cite{Burgess:2004kv} for an early attempt at joining both stories together).

This type of scenario could arise, for example, in a string model where the
Standard Model arose from branes at a singularity which are separated in the
bulk from the throat where inflation occurs.  In an
intersecting brane model scenario, however, it is not clear
that one can describe the full Standard Model by branes confined to a throat.
In most known constructions in Type IIB, one is forced to use magnetized D9-branes
(D9-branes with worldvolume magnetic fields turned on) as part of the visible sector
in order to generate the SM gauge group and matter content{\footnote{There are some
$U(4)\times Sp(2) \times Sp(2)$ models for which the SM sector contains no
magnetized D9-branes.  For these models, a D3-brane would only couple
gravitationally.  But for more general  constructions, this seems not to be the case.}}.
Since this D9-brane
wraps the full compact space, it cannot be separated from the inflaton sector.
The dual statement in IIA brane models is that the branes in question are D6-branes
wrapping 3-cycles on the compact space; such 3-cycles generically intersect each
other topologically, and so cannot be separated.

In this work, we describe a
scenario for $D$-term inflation \cite{Dterm} which seems to appear straightforwardly
in intersecting brane models (IBM).
In this scenario, the inflaton and waterfall fields appear as non-vectorlike (bifundamental)
matter arising from strings stretching between different hidden sector branes.

One of the basic difficulties in embedding $D$-term inflation in string theory is
that we require a flat direction (which is our inflaton), but at the same time
require that all tachyonic (``waterfall") directions which could lower the
inflationary potential are given positive mass.  In our context of Type IIA/B
IBM's, we naturally tend to find a large number of D-flat directions.  One can
see this simply by noting that if we have $N$ $U(1)$ brane stacks then we have
$N$ D-term equations.  But since each of these brane stacks generically has non-zero
topological intersection with the others, we have ${\cal O}(N^2)$ scalar fields
in the D-term equations.

Our basic idea will be to separate out one $U(1)$, whose $D$-term
potential will give us the inflationary potential \be V_{\rm{inf}}
^D = g^2(|\phi_+|^2 - |\phi_-|^2 - \xi)^2\; , \ee where $\xi$ is a
Fayet-Illioupoulos term that will give us the necessary vacuum
energy during inflation when $\phi_\pm = 0$. The remaining $D$-terms
will form \be V_{\rm{rest}} ^D  = g_j^2\left(|S|^2 + \sum_i q_i
|\rho_i|^2 -\xi_j\right)^2 + \cdots \, \ee where the dots represents
other terms of the same form as the first one, $\rho_i$ are various
chiral matter with charge $q_i$ under this particular $U(1)$ gauge
group. We now seek a flat direction for $V_{\rm{rest}} ^D$ which
lifts tachyonic directions for $V_{\rm{inf}} ^D$ via Yukawa
couplings of the type $\lambda S\phi_+\phi_-$. Hence we can give a
large initial vev to $S$ ensuring that the mass squared of $\phi_\pm$
are positive.  Given enough $\rho_i$ fields we can ensure that there
exists a flat direction for $V_{\rm{rest}} ^D$ spanned by the fields $S$ and
$\rho_i$. Our flat direction can thus naturally
act as the inflaton, which will tend to roll to the origin as a
result of the one-loop Coleman-Weinberg type potential.  As the
inflaton rolls to the origin, eventually the Yukawa couplings will
no longer lift the waterfall fields sufficiently. They will become
tachyonic (in our case $\phi_+$ will be the waterfall field) and
cause inflation to end (as usual in $D$-term inflation).  Of course,
one must worry that Yukawa couplings could also lift the flat
directions, but we will see that with some mild fine-tunings we can
ensure that the superpotential contributions are Planck-suppressed.
What aids us in our quest is gauge invariance; the fact that the
inflaton and waterfall fields are charged under hidden-sector groups
limits the types of couplings which we must worry about.

This type of slow-roll inflation generically predicts a near gaussian red spectra ($n \sim 0.98$) with an
inflationary scale of order of the GUT scale. The inflaton vev is always lower than the Planck scale
(which is necessary for a consistent compactification) and therefore it predicts a negligible amount of
tensor fluctuations. While the simplest model predicts the formation of cosmic strings with a tension
that is in conflict with the current bounds, we argue that simple extension can make the cosmic string unstable.

Since this scenario is set in the context of IBM's, it offers a rich
phenomenology and it is possible to tie in
various phenomenological processes with the inflationary phase. In particular, we show that this scenario
naturally admits a type of hidden sector baryogenesis, fed by tachyonic
preheating at the end of inflation.
We also note that it is possible to generate the neutrino masses via a see-saw mechanism at
the end of inflation. As such, we believe our model is very minimalist as we use the phase transition at the
end of inflation for multiple purposes.

Although our scenario seems to arise most naturally in intersecting brane
models, it can be studied purely from a low-energy effective field
theory point of view.   In fact, we do not
present a specific implementation of this model into string theory in
this paper. Instead we present the general conditions that a string
theory vacua will have to satisfy in order not to spoil the
inflationary epoch or the reheating era. We also discuss how moduli stabilization
and the uplifting to a de Sitter vacua might affect our inflationary scenario.  Since our potential
is coming from the D-term instead of the F-term we expect that we are less sensitive to these issues
but in the absence of a complete string theory model, we cannot completely elucidate these questions.

The paper is organized as follows. We start with a brief review of
intersecting brane model constructions in section 2.  We then present a method to achieve
 D-term inflation in the hidden sector in section 3 with particular
attention to various fine-tuning issues. Finally, we show how one can use our
high scale inflationary physics to solve other beyond standard model phenomenology
which might be happening at that scale.
In particular we describe how to achieve hidden
sector baryogenesis with tachyonic preheating in section 4 and how one could generate
the neutrino masses in section 5. Section 6 contains our
conclusion.

\section{Intersecting Brane Models}

We begin with a compactification of Type IIA/B string theory
to 4 space-time dimensions with $N=1$ supersymmetry and with all
closed-string moduli fixed.  There is considerable evidence suggesting
that this objective can be achieved in both
Type IIA and Type IIB by compactifying on an orientifolded
Calabi-Yau 3-fold and turning various RR, NSNS and/or geometric fluxes.
We will not rely on any specific mechanism for fixing the moduli, however,
and will simply assume that we are left with a 4D theory with $N=1$ SUSY
and no moduli.

In known compactifications of this form, one is typically required to
add in spacetime-filling D-branes in order to cancel RR tadpoles (essentially,
to cancel spacetime-filling charges inherent to the compactification, so that
there are no net spacetime-filling charges which would violate Gauss' Law).
In Type IIA, these branes are D6-branes which fill spacetime and wrap 3-cycles
on the 6D compactification manifold.  The Type IIB version is related to this
by T-duality (equivalently, mirror symmetry on the CY 3-fold), in which the
branes are most generally D9-branes with magnetic fluxes turned on, which generate
lower brane charge (and they appear commonly in known IBM constructions).  We will focus on the Type IIA description, as the spectrum
has a more geometric interpretation in that case.

The open strings which begin and end on these D-branes will generate a low-energy
gauge theory with chiral matter.
Strings which begin and end on a single stack of $N$ branes will provide the degrees of
freedom of a $U(N)$ gauge multiplet\footnote{$SO$ and $Sp$ gauge groups are also
possible if the brane stack lies on an orientifold plane.  In several examples in the
literature, the gauge group rank $N$
is related to the number of branes in the brane stack by a non-trivial factor.  This
results from constructions in which the Calabi-Yau 3-fold arises in an orbifold limit,
where some gauge degrees of freedom are projected out.  At our level of generality we
need not worry about this subtlety; in all relevant examples there will exist a number
of branes in a stack which would yield $U(N)$ gauge group for any integer $N$.}.
Moreover, for any two branes $a$ and $b$
with gauge groups $G_{a,b}$, there will be chiral multiplets transforming in the
bifundamental arising from strings stretching between the two branes.  The net number
of such multiplets is given by the topological intersection number $I_{ab}$.  As
these branes wrap 3-cycles on the compact 6D space, any two branes will generically
have non-zero topological intersection number, and thus will have net chiral matter
transforming in the bifundamental.

The intersecting brane models which are relevant for phenomenology have a set
of ``visible" sector branes on which live a gauge theory with
MSSM gauge group (or a basic extension thereof) and with SM chiral matter content
arising from strings stretching between different visible sector branes.  There are
several known constructions of models of this kind in both Type IIA and Type IIB.
In general, these models will have additional branes beyond those which yield the
SM gauge theory, the so-called ``hidden sector."  These branes appear generically
because of the need to cancel the RR-tadpoles generated by the compactification;
there is no reason why ``visible sector" branes alone should generate precisely the
charges needed to cancel these tadpoles, and in all known constructions additional
branes are required.  It is this hidden sector which will be relevant for the
D-term inflation model which we discuss here.

One might worry that the appearance of such generic chiral matter would create anomalies
which would destroy the consistency of the compactification.  In fact, the cancellation
of RR and K-theory tadpoles is both necessary and sufficient to cancel
all cubic anomalies\cite{MarchesanoBuznego:2003hp}.
Mixed $U(1)$ anomalies can still arise and are generally cancelled by the Green-Schwarz
mechanism.  These mixed anomalies will play an important role in our proposed model of
 baryogenesis.

There are many known IBMs on different orientifolded CY 3-folds (most commonly,
$T^6 / Z_2 \times Z_2$).  Our interest is not in any particular model; indeed, if
intersecting brane models do provide a realistic description of the world, then the
relevant IBM is likely not among the models already constructed.
Thus, our interest is in a scenario for inflation
which can occur in a large class of such models.  In particular, we will not
constrain our IBMs to match any particular scheme for moduli stabilization, or any
preferred low-energy model of phenomenology, provided they can still match data.
As we
have mentioned, these IBM constructions naturally arise with $N=1$ supersymmetry,
due to their origin as a string compactification.  This supersymmetry is also relied
on in some closed-string moduli stabilization schemes
such as KKLT \cite{GKPKKLT} to assure control of the
potential (though it is by no means clear that supersymmetry is necessarily required in
all constructions
to give good control over the potential \cite{eva}).
However, the breaking of supersymmetry and the applicability of
supersymmetry to the hierarchy problem is an open question.  One can certainly attempt
to implement familiar types of supersymmetry breaking mediated to the visible sector
in such a way as to generate a Higgs mass which is naturally small.  But other options
have also drawn attention.
Attempts to  match LEP data to the MSSM have led to the so-called little hierarchy
problem, which suggests to some that perhaps an alternative to low-energy SUSY is
required, and various mechanisms (for example, the little Higgs, twin Higgs, environmental
selection, ``stringy" naturalness, etc.\cite{nonlowscale}) have
been proposed to deal with the hierarchy without supersymmetry.
We will be agnostic about all such claims; we demand supersymmetry at the string and
moduli scale for control of the potential, but will admit any supersymmetry breaking
which arises below that scale.

\section{D-term Inflation in the Hidden Sector}

There has been much recent activity in working out models of the early
universe and, in particular, models of inflation in string theory.  Most of the
recent activity has been set in the context of type IIB since, in this case, explicit
constructions of de Sitter vacua with all moduli stabilized have been worked out \cite{kklmmt}.
In particular, a string theory implementation of D-term inflation has already been
proposed in this context \cite{d3d7}. In these models the inflaton field is the distance
between a D3 and a D7 and the vacuum energy is provided by fluxes on the D7.  From a
four dimensional view point, these fluxes generates a field dependent FI term. As the
D3 brane get closer to the D7 brane, the open string stretching between them becomes tachyonic and
inflation ends.

One expects there to exist Type IIA constructions which are mirror to the uplifted Type IIB
constructions.  However, such constructions could be much more complicated to realize explicitly in
Type IIA language.
Uplifting vacua in type IIA string theory has proved to be a rather technical
issue and no explicit construction exists at the moment \cite{Kallosh:2006fm}.
It appears that one might need a combination of D-term \cite{BKQ} (with the caveats of \cite{Choi})
and F-term
breaking \cite{GKPKKLT} to uplift the vacua (as for
example in \cite{Giovanni1}) and it remains to be shown
that this can be accomplished in a specific model where all the corrections are under control.
But we will not need to rely on the details of the moduli stabilization or
of the uplifting of the final vacuum to
de Sitter,\footnote{There has been some recent work discussing
the effect of moduli stabilization on D-term inflation in \cite{Brax:2006yq} (this is
in a specific D-term uplifted model \cite{anna}). F-term inflation in this model
was also discussed in \cite{deCarlos:2007dp}. }
only on their existence.  For this purpose, one may consider Type IIA just as well as
Type IIB as an arena for inflation in intersecting brane models.

We can now describe the general Type II intersecting brane models we are considering.
Although we will count non-vectorlike matter representations by the topological
intersections of branes, which is the appropriate geometric picture in Type IIA,
similar formulae and results apply in Type IIB via mirror symmetry.
We are interested in getting inflation in the open string sector from the various
non-vectorlike bifundamental matter fields that are ubiquitous in
IBMs (we denote these fields $\chi_i$
in the following).  We suppose that all the adjoint and other vectorlike
open string matter are also fixed around
the string scale and for simplicity we do not consider the symmetric and antisymmetric matter
that arises from strings stretching to the orientifold plane.  Since each $\chi$ is charged
under two gauge groups (both of which contains a U(1) in general) we will get various
Fayet-Illioupoulos terms in addition to the usual non-abelian piece. Given $N$ stacks of branes
with various gauge symmetries $U(n_1) \times U(n_2) \cdots U(n_N)$, we get
\be
V_D \sim \sum_{j=1}^{N} g_j^2 \left(\sum_i q_{i,j} |\chi_i|^2 -\xi_j\right)^2 + V_{\rm{NA}}\; ,
\ee
where $\xi_j$ are the various FI terms, $q_{i,j}$ are the charges of the $\chi_i$ fields
under the gauge group $j$ (this is non-zero for two gauge groups only) and $V_{\rm{NA}}$ is
the non-abelian contribution to the $D$-term. From gauge invariance the only renormalizable terms
in the superpotential will come from Yukawa couplings of the following type
\be
W \sim \sum_{i,j,k} \lambda_{ijk} \chi_i\chi_j\chi_k + \mathcal{O}(M_p^{-n})\; ,
\ee
where the sum runs over all combinations that are allowed by symmetry.  Assuming that all
the $\chi_i$ fields have subplanckian value and canonical kinetic term, we get the following
F-term potential
\be
V_F \sim \sum_{i,j}  \lambda_{i,j}^2 |\chi_i|^2|\chi_j|^2 + V_{F,\rm{other}} + \mathcal{O}(M_p^{-2n})\; ,
\ee
where $V_{F,\rm{other}}$ is a constant potential depending on
the vevs of all the other stabilized moduli.
It is fixed by matching to the correct cosmological constant at the end of inflation.
Note that this is all very schematic (we will work out an explicit
example in the next subsection) but this is enough to present the generic idea.  At
first all the $\chi$ fields are on the same footing
and there is no favored clock or inflaton field.

We pick out one of the $D$-terms (assume it is a $U(1)$ for the moment, though that is not essential)
to act as the inflationary potential.  We can then classify our $\chi$ fields into three different
sets:
fields with charge $+1$ and $-1$ under
that particular gauge group ($\phi_{\pm,i}$)
and fields $S_i$ that are neutral under that gauge group.  For
a convenient choice of the sign
of $\xi$, the fields which charge $+1$ are the potential waterfall fields.
By giving a large enough initial vev to some of the $S_i$ we can make sure
that all the $\phi_{\pm,i}$
are massive through their Yukawa couplings.  Since we have a large space
of flat directions for the
remaining $D$-terms (excluding our inflationary term), we can simply move out along one of those directions,
and use the resulting vevs to give mass to the waterfall fields.
We then have a very flat potential that gives rise to the usual $D$-term inflation
as the $S_i$ gradually roll
back to the origin along the flat direction, due to one-loop logarithmic
corrections.  As the $S_i$ fields roll back
towards the origin, the effective $mass^2$ given to the waterfall fields
(here the $\phi_{+,i}$) will drop and
one of them will eventually become tachyonic and condense, ending inflation.
The $\phi_{-,i}$ are spectators
fields and they are massive throughout the whole process.

We need to make sure that our $D$-flat direction is not lifted by Yukawa couplings.  Any flat
direction must involve turning on a set of fields such that the contribution of any field to a $D$-term is
compensated by the contribution of other fields with the opposite sign charge to the same $D$-term.  For example,
if one drew the quiver diagram corresponding to the brane gauge theory, any closed oriented polygon would
correspond to a possible flat direction.  The only Yukawa coupling permitted by
gauge invariance would be $W = {\lambda \over M_{pl} ^{n-3}} \phi_1 ... \phi_n$, where $\phi$ are the fields
forming the closed oriented polygon and $n$ was the number of sides.  Only for a triangle would the
Yukawa coupling be renormalizable; for higher $n$ the Yukawa couplings would be highly suppressed,
yielding a very flat potential.  To avoid the appearance of ``triangles" involving fields
of the inflaton direction requires only the tuning of
the sign of a few topological intersection numbers.  Let us illustrate this mechanism
in more details by working out the simplest implementation.

\subsection{The Simplest Model}
We will consider six stacks of
hidden sector branes $a$, $b$, $c$, $d$, $e$ and $f$.  These branes each fill
spacetime and wrap different 3-cycles on the compact 6D space.
The topological intersection numbers of our toy model are
$I_{ab}=I_{bc}=I_{ca}=I_{be}=I_{ef}=I_{fa}=I_{db}=I_{ad}=1$,
with all other intersection numbers being zero.  We can thus readily
see that all cubic anomalies cancel.
Each brane stack yields a $U(1)_{a,b,c,d,e,f}$ gauge group.
We denote by $S$, $\phi_+$ and $\phi_-$ the chiral multiplets stretching
between $a$ and $b$, $b$ and $c$, and $c$ and $a$ respectively (see Fig. 1).
\begin{figure}[ht]
\centering
\includegraphics[width=10cm]{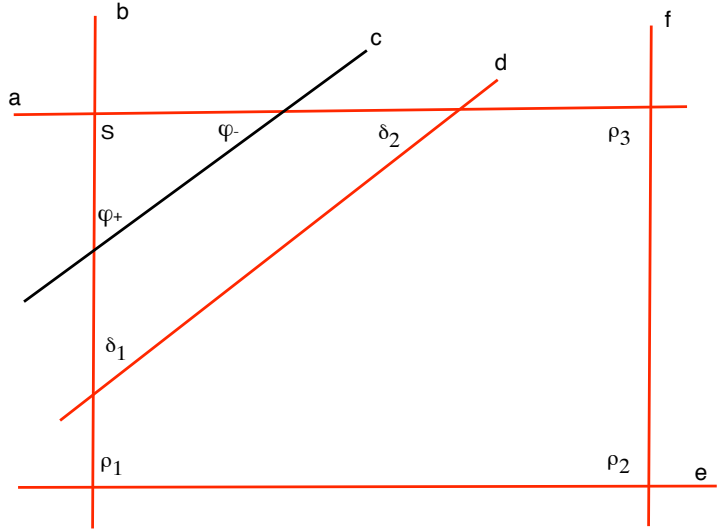}
\caption{This shows a setup of intersecting branes in type IIA that can yield $D$-term inflation.
We have a total of 6 branes. Brane $c$ is our inflationary brane.
Branes $a$ and $b$ give rises to the necessary Yukawa coupling between the
inflaton and the waterfall field.  Branes $e$ and $f$ are there to ensure that
the $D$-term potential has a flat direction.  This flat direction is broken by the
$F$-term potential but this is planckian suppressed. Finally,
brane $d$ is necessary for the anomalies to cancel, it plays no other role
during the inflationary era. $S$, $\phi_\pm$, $\delta_{1,2}$ and $\rho_{1,2,3}$ are the various
bifundamental fields at each intersections. Also note that this picture is a
two dimensional depiction of a configuration that actually lives in six curved dimensions.}
\end{figure}
The
$U(1)$ gauge group living on each brane stack
will contribute a $D$-term to the
Lagrangian.  We thus obtain the $D$-term potential
\bea\label{fullDterm}
V_D &=& g_c ^2 (|\phi_+| ^2 -|\phi_-| ^2 -\xi_c)^2
+g_a ^2 (|\phi_-| ^2 -|\delta_2 |^2 -|S| ^2 +|\rho_3|^2-\xi_a)^2
\nonumber\\
&\,&+g_b ^2 (|S| ^2 +|\delta_1|^2 -|\phi_+| ^2 -|\rho_1|^2 -\xi_b)^2
+ g_d ^2 (|\delta_2|^2 -|\delta_1|^2 -\xi_d)^2
\nonumber\\
&\,&+g_e ^2 (|\rho_1|^2 -|\rho_2|^2 -\xi_e)^2
+g_f ^2 (|\rho_2|^2 -|\rho_3|^2 -\xi_f)^2\; ,
\eea
where $g_{a,b,c,d,e,f}$ are the gauge couplings of the gauge theories
on the three branes, and $\xi_{a,b,c,d,e,f}$ are the Fayet-Iliopoulos
terms.  These FI-terms are set by the angles between the various
branes, and thus are functions of the topological wrapping numbers
of the branes and of the complex structure moduli (in Type IIA;
K\"ahler moduli in Type IIB) (see e.g. \cite{Giovanni2} for a recent discussion).
Since we have not dealt with the details of how the moduli are fixed,
we cannot solve for these moduli.  But there are known mechanisms which
are believed to provide a very large number of vacua, each of which
has different values for the vevs of the moduli.  Relying on this type
of statistics, we will treat the FI-terms as tunable parameters (the
tuning would occur by selecting a particular metastable vacuum to
expand around).

For simplicity we
set $\xi_d=\delta_1=\delta_2=0$.  We choose the $c$ brane stack as our
inflationary potential, and we have
\bea
V_{\rm{inf}} ^D &=& g_c ^2 (|\phi_+| ^2 -|\phi_-| ^2 -\xi_c)^2\; ,
\nonumber\\
V_{\rm{rest}} ^D &=&
g_a ^2 (|\phi_-| ^2 -|S| ^2 +|\rho_3|^2-\xi_a)^2
+g_b ^2 (|S| ^2 -|\phi_+| ^2 -|\rho_1|^2 -\xi_b)^2
\nonumber\\
&\,&+g_e ^2 (|\rho_1|^2 -|\rho_2|^2 -\xi_e)^2
+g_f ^2 (|\rho_2|^2 -|\rho_3|^2 -\xi_f)^2\, .
\eea
The flat direction is the path in field space along which the variation
$\delta(|S|^2) = \delta(|\rho_{1,2,3}|^2) = r^2$ and
 $V_{\rm{rest}} ^D$ is constant \footnote{One might worry that, along
this flat direction, $V_{rest} ^D$ will be constant but non-zero.
This would not change our picture provided the potential is small.  But
in general, there will be additional fields arising in symmetric and
anti-symmetric representations, as well as fields transforming in the
fundamental of two gauge groups.  These fields come from strings stretching
between orientifold images, and we have omitted such terms for simplicity.  But
vevs for these fields will generally permit $V_{\rm{rest}} ^D=0$.}. This flat direction
constitutes our inflaton.

A Yukawa coupling will
arise from  a worldsheet instanton stretching over the triangle generated by the
three intersecting branes $a$, $b$ and $c$ or $d$:
\be
W_{\rm{yuk}} = \lambda_\phi S \phi_+ \phi_- + \lambda_\delta S\delta_1\delta_2\; .
\ee
Due to gauge invariance, these are the only two renormalizable terms in the
superpotential containing these fields.
The only Yukawa coupling relevant to the flatness of our inflaton direction
is
\be\label{square}
W_{2} = {\lambda_2 \over M_{pl}} S \rho_1 \rho_2 \rho_3\; .
\ee
Note that this coupling is Planck-suppressed because it arises from
a ``square" in the quiver diagram, rather than a triangle.  Higher
suppressions are easily allowed by simply choosing more hidden sectors branes
such that the flat direction is described by a polygon with more sides.
Since the uplifting of the flat directions can be suppressed sufficiently,
we will simply drop this term for the moment.  And since we are still on a flat-direction
of $V_{\rm{rest}} ^D $, we will  set this term to zero as well.
We will take canonical K\"ahler
potential for these open string fields
\footnote{The exact form of the K\"ahler potential for non-vectorlike
open string fields is not known.  The leading moduli dependence of this
K\"ahler potential was computed in \cite{Lust:2004cx}, and when
$\xi \ll m_s ^2$ the dependence on the size moduli is small.  Note that higher order corrections
to the K\"ahler potential will not affect the $D$-flatness 
of $V_{\rm{rest}}^D$ but it will in general lead to
corrections to the mass terms of $\phi_\pm$.}
Neglecting planckian corrections and setting $\delta_{1,2} = 0$, the F-term
potential in this model is:
\be
V_F = \left( \lambda_\phi^2(|\phi_+|^2 |\phi_-|^2 + |S|^2 (|\phi_+|^2 + |\phi_-|^2)) + V_{F,\rm{other}}\right)
\ee
where $V_{F,\rm{other}}$ is the constant potential from all other F-term SUSY-breaking
sectors. In order to have $D$-term inflation we need $F \ll D$.  Although this is 
a working assumption throughout this paper, it could be relaxed and one could have an $F$ and $D$ term of the same order 
(however, crucial details of the inflaton potential would change in this case). So we will assume the $F$-term
potential is subdominant and we will drop it for the rest of this discussion (note
that one cannot have a strictly vanishing $F$-term when the $D$-term is non-vanishing in supergravity).

We will work in the regime where the inflaton vev is always smaller than
the string/Planck-scale, as well as below the masses of the moduli.
As such, we are justified in ignoring
gravitational corrections to the K\"ahler potential, and in using an
effective field theory approach with excited strings and moduli integrated out.
It is possible that these constraints can be relaxed consistently with
inflation, but we will assume them in order to obtain a simple and
standard framework.
Our scenario for inflation is thus very ``field theoretic," and in fact can
be implemented in effective field theory without necessarily relying on its string
origin and motivation.
Note that it is possible to have some moduli
unstable during inflation as long as their effects on the inflationary predictions are under
control and that the moduli
stabilize at a phenomenologically viable value at the end of inflation.

Note that, a priori, nothing picked out $S$ as the
inflaton.  It is the initial conditions and the existence of the appropriate
flat directions which
picks out the appropriate field to act as the inflaton. Also we should
emphasize that this model differs from the originally proposed $D$-term inflation
model \cite{Dterm} since the inflaton field is in a non-vectorlike representation.
The tree-level potential is
\be
V_{\rm{tree}}= g^2 (|\phi_+| ^2 -|\phi_-| ^2 -\xi)^2
+\lambda_\phi^2 (|\phi_+|^2 |\phi_-|^2 + |\phi_+|^2 |S|^2
+|S|^2 |\phi_-|^2)\; .
\ee
The global supersymmetric minimum for this potential occurs at
$|\phi_+|^2 =\xi$, $\phi_- =S =0$.  However, for
$|S| > S_C = {g\sqrt{2\xi} \over \lambda}$, the non-vanishing
minimum of the
potential is at $\phi_+ = \phi_- =0$.  This potential  is thus of
the hybrid form, with $S$ the inflaton field and $\phi_+$ the
waterfall field.  But one should remember that the actual inflationary
path involves more than just the field $S$.  In some sense we may
thus regard this as a multi-inflaton model, where  $S$ is the inflaton
which dictates when the waterfall field starts rolling.

For $|S|> S_C$, we can thus set $\phi_{\pm}=0$ and write an effective
potential for $S$.  This potential  will have a one-loop correction
of Coleman-Weinberg form, yielding
\begin{align}\label{potential}
V_{eff} (S) & = g_c^2 \xi^2 \left[1+ { g_c^2 \over 8 \pi^2}
\log \left({\lambda_\phi^2 |S|^2 \over \Lambda^2}\right) + { g_d^2 \over 8 \pi^2}
\log \left({\lambda_\delta^2 |S|^2 \over \Lambda^2}\right)\right]\; ,\nonumber \\
 &\approx g^2 \xi^2 \left[1+ {g^2 \over 4 \pi^2}
\log \left({\lambda^2 |S|^2 \over \Lambda^2}\right)\right]\; ,
\end{align}
where we have taken $g_c\sim g_d \equiv g$ and $\lambda_\phi \sim \lambda_\delta \equiv \lambda$
in the second line.
Note that the one-loop contribution is dominated by the $S$ dependence.  This
is the case because the Coleman-Weinberg contribution comes from the splitting
of boson and fermion masses.  But if we are on a flat direction of $V_{\rm{rest}} ^D$,
then the boson masses are not split by these $D$-terms (the FI contribution is
cancelled by the contribution from the vevs of the fields).  The only contribution
to boson/fermion mass splitting comes from $V_{\rm{inf}} ^D$, and only $S$ couples
to these fields via the Yukawa coupling to generate the fermion mass.
So for practical purposes now, $S$ is acting as our inflaton.
Note also that any pair of fields that couples to $S$ through a Yukawa coupling
will contribute to the Coleman-Weinberg potential \footnote{Note that in the literature, one usually
has a factor $\frac12$ in front of $g^2\xi^2$.  Hence our $g_{\rm{our}} = \sqrt{2} g_{\rm{other}}$.}.
In our case we have  two pairs $\phi_\pm$ and $\delta_{1,2}$.

Although this standard form of the potential
has been widely discussed in cosmology literature \cite{Lyth:1998xn}, it is not
trivial to obtain this form in string theory.  For example, one
typically
avoids a large tree-level mass for the inflaton by imposing a discrete
symmetry which protects it; it is more difficult to find such a
symmetry in string construction.  In this intersecting brane model
construction, the inflaton appears as an open string field in a
non-vectorlike representation and hidden sector gauge
invariance protects it from a tree-level mass.

When $|S| =S_c$, the waterfall field $\phi_+$ becomes
tachyonic and begins to condense.  As usual in hybrid inflation,
this condensation brings a rapid end to inflation.  In this case,
the condensation of $\phi_+$ field causes brane recombination, in
which the brane stacks $b$ and $c$ deform and bind to each other
in order to minimize their energy.  In general one has
to worry about
whether or not the effective field theory approximation is valid for
large values of the Fayet-Iliopoulos term.  But we are assuming the
limit $\xi \ll m_S$, in which case the field theory limit should be
valid.

Finally, note that this model predicts the formation of stable cosmic strings at the end
of inflation from the spontaneous symmetry breaking of the $U(1)$. As we will see, the
experimental bound on the  cosmic string is the biggest phenomenological constraint on
$D$-term inflation although a simple and natural extension of the model can easily solve
the problem.

\subsection{Phenomenological Status of D-term Inflation}

$D$-term inflation is a fairly robust model that has been extensively studied in the literature
(see \cite{urRehman:2006hu, Lyth:2007qh} and references therein for some detailed
analysis). In this section, we remind the reader of the simplest story. The $D$-term
inflationary potential is
\be
V = V_0 \left(1+ \alpha \ln \frac{\lambda s^2}{\Lambda^2}\right)\; ,
\ee
where the height of the potential is set by the FI term $V_0 = g^2\xi^2$, $\alpha = g^2/4\pi^2$
is the loop factor and we have defined $s \equiv |S|$. In the slow-roll regime, we can
solve for the inflaton field in terms of number of efolds $N$. Using $-Hdt = dN$, one has
\bea
3H\dot s =  -\frac{\partial V}{\partial s}\; ,\nonumber\\
3H^2\frac{\partial s}{\partial N} =  \frac{2V_0 \alpha}{s}\; ,
\eea
with can be solved approximately for constant
$H \approx \sqrt{\frac{V_0}{3M_p^2}}\sim \frac{g\xi}{\sqrt{3}M_p}$
\be
\frac{s^2 - s_0^2}{2} = \frac{2 V_0\alpha}{3H^2} N\; .
\ee
The term on the right side is just proportional to $g^2 M_p^2$.  Given a $g^2$ that is not too small,
this term will in general dominate over the $s_0^2$ and this is the limit
that we are going to work in \cite{Lyth:1997pf}.
Note that in the opposite limit where $s^2 \approx s^2_0 \sim |S_c|^2$,
we can still match the data although the observables will now depend on $\lambda$.
We require that the initial value of $s$ (which minimally corresponds to 60 efolds) satisfy
the following hierarchy of scales for the effective field theory approach to be valid
\begin{align}
s^2 &\ll m_\Phi^2 \lsim m_{\rm{str}}^2 \ll M_p^2 \; ,
\end{align}
where $m_\Phi$ is the mass scale of the string moduli and $m_{\rm{str}}$ is the string scale.
We note here that we expect a spread in the masses of the moduli (see for e.g \cite{Easther:2005zr}
where they study the spectrum of moduli masses via random matrix techniques).  In order for inflation to go
through unmodified, we probably need the value of the inflaton to be lower than most masses
of the moduli
which might be a stronger requirement than taking it to be smaller than string scale. Supposing
that we have
no more than 60 efolds, we get that the initial vev of $s$
is of order $s^2 \sim 6 g^2 M_P^2$.
One could potentially satisfy the needed hierarchy of scale with $g^2 \sim 10^{-2}$ or $10^{-3}$.

With subplanckian vev for $s$, we will have negligible tensor/scalar ratio $r$. Since this is
slow-roll, we expect
negligible non-gaussianities and a negligible running of the spectral index.   Hence the main
three observables are the
power spectrum, the spectral index and the cosmic string tension:
\bea\label{observables}
P_{\mathcal{R}}  & = & \left. \frac{1}{24\pi^2 M_p^4}\frac{V}{\epsilon} \right|_{k=aH} =
\frac{N}{3} \frac{\xi^2}{M_p^4}\; ,\\
n &\approx &1+2\eta = 1-\frac{1}{N}\; ,\\
G\mu & \sim &\frac{\xi}{M_p^2}\; .
\eea
The first two are evaluated 55 efolds before the end of inflation and the cosmic string tension
comes from the end of
inflation when the waterfall field condenses.  First, we find that the spectral index is $0.98$
which is within $\sim 2\sigma$ range
of WMAP3 year data \cite{Spergel:2006hy}.  The power
spectrum is measured to be $P_{\mathcal{R}}  \sim 10^{-9}$ from
which we get that the scale of inflation
is of order of the GUT scale
\bea
\xi \sim 10^{-5} M_p^2 \sim (10^{15} \gev)^2\; ,
\eea
which gives a cosmic string tension of $G\mu \sim 10^{-5} $ which is well in excess of the current
experimental bounds $G\mu \lsim 2\times 10^{-7}$ \cite{Wyman:2005tu}.

In this intersecting brane model setup, one can naturally
avoid the production of long-lived cosmic strings entirely 
by taking one of the gauge groups to be
non-abelian.   Alternatively, one can also avoid the production of
stable cosmic strings by having multiple waterfall fields $\phi_+ ^i$
in the bifundamental (this will occur in our toy model if $I_{bc} >1$).
In either situation, the vacuum manifold will have trivial $\Pi_1$ and
the produced cosmic strings will be either of the electroweak type for a local non-abelian
symmetry or semilocal for a global non-abelian symmetry \cite{semilocal}. Both 
type of strings are generically
unstable thus permitting them to decay and avoid any conflict with
observation \cite{Urrestilla:2004eh}.
Unfortunately, this resolution of the cosmic string problem will also remove
any possible observable signatures of cosmic strings.  It is argued that 
there is a region of the parameter space for which a complete fit of the model 
including the contribution of the cosmic strings to the power spectrum is 
possible \cite{Rocher:2004my}.  In this case it would then be possible to retain an 
observable cosmic string signature
which is consistent with current experiments.

It is interesting to note that, because generically these hidden sector
branes will have non-trivial topological intersection with the visible sector
branes, the Standard Model can be heated directly by open string degrees of
freedom.  In contrast, the current models of string inflation (for example, brane inflation)
heat the visible sector through gravitation degrees of freedom.  We will
see that this different feature has implications for baryogenesis.

Finally, we
can quantify how much fine-tuning is needed in order not to spoil the slow-roll conditions.
Although we have enough scalar fields to make
the $D$-term flat, we will still generate $F$-term contribution to the potential from the
quartic Yukawa superpotential.  Since the log grows very slowly for large $s$,
these contributions might ruin the slow-roll condition.
During inflation we have $\rho_i^2 + \xi_j \sim S^2$.  From the superpotential (\ref{square}), we
will generate a term $\sim \frac{\lambda_2 ^2}{M_p^2}S^6$ as well as lower dimensional terms. Note
that lower dimensional term are suppressed by factor of $\xi/M_p$ which we have assume are all
small ($\xi_c$ is the biggest one of the FI terms and it is itself much smaller than $M_p$) and
therefore the most dangerous term is $S^6$.

Now in general a term like $\frac{\lambda_n ^2 S^{2n+2}}{M_p^{2n-2}}$ will
contribute to the $\eta$ parameter (see \cite{Kolda:1998kc} for a discussion of the
importance of these type of terms in $D$-term inflation):
\be
\eta \sim \lambda_n ^2 (2n+2)(2n+1) \frac{S^{2n}}{V_0 M_p^{2n-4}}\; ,
\ee
taking $S^2 \propto 6g^2M_p^2$ at 55 efolds before the end of inflation and the definition
of the power spectrum (\ref{observables}). We find that requiring $\eta < 10^{-2}$ put the following
constraints on $g^2$ and $\lambda_n^2$.
\be\label{finetuning}
\lambda_n ^2 (g^2)^{n-1} < \frac{P_{\mathcal{R}} 10^{-2}}{6^n (2n+1)(2n+2)}\; .
\ee

In our simplest model ($n=2$), we get an $S^6$ term which requires a fine-tuning
\be
\lambda_2 ^2  g^2 < 10^{-13}\; .
\ee
This could be satisfied if we had $\lambda_2 \sim g \sim 10^{-3}$.
Note that the flat direction can easily arise from a polygon with more sides in our
intersecting brane setup.  In this case, the $F$-term will come with additional $M_{pl} ^2$
suppression which reduces this fine-tuning.  In any case, we expect gravitational
corrections to the K\"ahler potential to yield potential terms of the form
$\Delta V \sim ({|S|^2 \over M_{pl} ^2})^n\langle V_{F,\rm{other}} \rangle$.  For low-scale
supersymmetry breaking
we expect $\langle V_{F,\rm{other}} \rangle \ll V_0$ and these corrections will be
small.  For higher-scale $F$-term supersymmetry breaking, one would have to cancel these
corrections against the Yukawa terms. This would be an $\eta$ problem, but if the $F$-term is large
than this model can no longer be considered a purely $D$-term inflationary model.

We see from (\ref{finetuning}) that by
using a polygon with enough  sides ($n$= 5 should do) one can effectively eliminate the
needed fine-tuning.
Hence it is possible to brane engineer a flat direction.  It is also worth
noting that the $\lambda_2$ Yukawa coupling is generated by worldsheet instantons,
which generally are exponentially suppressed by $\exp(-A / l_s ^2)$, where $A$ is the area of
the string worldsheet stretching between the brane stacks forming
the polygon.  In the large volume
limit (where known moduli-stabilization schemes are under control), one would expect $A>1$ in
string units, implying that the coefficient $\lambda_2$ should be relatively small.

On the down side, $g^2$ should really
be replaced by $\sum g^2$ in the case where there are multiple Yukawa couplings between $S$ and other fields.
Hence the fine-tuning in equation (\ref{finetuning}) should be done on every one of the gauge
coupling constant that appears in the Coleman-Weinberg potential.  Remember that equation (\ref{finetuning})
was obtained assuming $g_c \sim g_d$ and so we need to fine-tune both coupling constants.

Finally, since this model is really a multiple fields inflationary scenario,
one might expect that isocurvature perturbations will be generated by the fluctuations
of the $\rho_i$ and these perturbations might modify our fit to the data.  These would
come from coupling between the $\rho_i$ and $S$ in the $F$-term and since we are arguing
that these should be small, we expect the isocurvature perturbations to be small as well.

\subsection{More complicated/generic setups}

This setup is perhaps the simplest version of our scenario which would arise in
intersecting brane models, but it is not necessarily the most generic.  One might
expect, for example,  that the various stacks could contain multiple
parallel branes (non-Abelian gauge groups) and that the various stacks could
have $|I_{xy}| >1$ (more than one multiplet in the various bifundamentals).  We
will see that these more generic setups provide new constraints, but also new
phenomenological features.

The first serious worry is the possibility of multiple waterfall fields for
the inflaton potential (the $D$-term potential associated with the $c$ brane
stack in our previous example). One source of new waterfall fields would
be additional branes stacks
with topological intersection with the
brane stack $c$.  Another source would be multiple topological intersections
(for example, suppose in our previous example that we had $I_{bc}>1$).  If there
exists a new brane stack $g$ with non-trivial topological intersection with $c$,
then the new $D$-term potential will be of the form
\be
V_D = g^2 (|\phi^i _+|^2 - |\phi^j_-|^2 \pm |\rho^k|^2 +...-\xi)^2
\ee
where $\rho$ is the scalar arising from strings stretching between
$g$ and $c$, and the superscripts index our choice of chiral multiplet (if we have
multiple chiral multiplets transforming in various bifundamentals).
The charge of $\rho$ is determined by the sign of the topological
intersection.  In this more generic setup, one sees that there can be multiple
waterfall fields which could condense, causing $V_{inf} ^D$ to go to zero.  For inflation
to occur, it is necessary for every waterfall field to get a mass which stays positive
until the inflaton reaches its critical value (and inflation ends).
But in general this should not be a problem.  Recall that if we have $N$ brane
stacks generating $N$ $D$-term equations, we generally have ${\cal O}(N^2)$ scalars
which we can use to find flat directions.  Requiring those flat-directions to lift
the ${\cal O}(N)$ waterfall fields of $V_{\rm{inf}} ^D$ should still give us
${\cal O}(N^2)$ acceptable inflaton directions (modulo the fine-tuning of the signs
of intersections numbers to obtain Planck-suppression of Yukawa
couplings; for a flat direction given by an $n$-sided polygon, this means
a choice of ${\cal O}(n^2)$ signs).

Another subtlety worth noting arises if one of the gauge groups is non-abelian.
In this case there will be non-Abelian $D$-terms
of the form
\be
V_r ^D = (\sum_i \phi_i ^{\dagger} T_r \phi_i )^2
\ee
where $T$ is a non-diagonal generator of the gauge group and $\phi_i$ is any
scalar charged under the non-abelian symmetry (there is no FI-term
for these terms).  If $G_c$ is a non-Abelian group, then one must worry if the
presence of the new $D$-term equation will prevent the full $D$-term potential
from dropping to zero when $\phi_+$ condenses.  Indeed, if $\phi_+$ is the only
waterfall field then $V_D ^{total}$ will
reach an ${\cal O}(1)$ fraction of the original value of $V_0$ (due to the non-zero
final value of $V_r ^D$).  In this case, the waterfall
will still end inflation but we would now need to be sure that the cosmological constant
at the end of the
waterfall is small to match
observation.  This would require a cancellation between $V_D$ (which is an ${\cal O}(1)$
fraction of the inflation scale), and $V_F$, which in turn presents an $\eta$-problem
which must be resolved by fine-tuning.
Since we would have $D$-terms
of order $V_0$, we would in this case be led to a model of high-scale supersymmetry
breaking.  But it is important
to note that this is only true in the simple example where there is only one waterfall
field.  If for example, we had $I_{bc}=2$, we would have two waterfall doublet
fields $\phi_+ ^{1,2}$.
In this case, one can see that by condensing the upper component of the $\phi_+ ^1$
and the lower component of $\phi_- ^2$, one can set to zero all of the $D$-terms
associated with brane stack $c$ (furthermore, if $I_{ca}=2$, then the inflaton $S$ alone
can give mass to all the waterfall fields via generic Yukawa couplings).
Another way of phrasing this would be to note that although non-Abelian
$D$-terms will give us more constraints than our original ${\cal O}(N)$,
higher topological intersection numbers can give us more fields than
our original ${\cal O}(N^2)$, so that we can still find sufficiently
many flat directions.

Finally, we briefly discuss a subtlety mentioned in \cite{Binetruy:2004hh}.
As they noted, it is potentially difficult to maintain low-energy supersymmetry
while simultaneously fixing the FI-terms at a high-scale (i.e., fixing the
closed string moduli on which the FI-terms depend at a high scale).  If
the $U(1)$ associated to the $D$-term is anomalous (with the anomaly fixed
by the Green-Schwarz mechanism), then the vector boson becomes massive by
eating the axion.  However, the axion itself is in the same multiplet as the
K\"ahler moduli, which are assumed fixed at a high scale.  If supersymmetry
is preserved, then this would imply that the vector multiplet becomes
massive at a high-scale, in which case one should not see the $D$-term equation
at low energies.  One might worry that our scenario would face this difficulty,
as the appearance of so many non-vectorlike representations would imply
that many $U(1)$'s have some mixed anomalies.

Of course, as noted in \cite{Binetruy:2004hh}, this $D$-term picture can
be made consistent by some amount of $F$-term supersymmetry breaking, and
this amount can be made very small.  This picture is completely consistent
with our scenario.  But one should remember that in our scenario, it is
in fact possible to generate many non-anomalous $U(1)$'s.  One can see this
simply by noting that the number of anomalous $U(1)$'s with anomalies fixed
by the Green-Schwarz mechanism is limited by the number of axions present
(basically, by the number of size moduli in Type IIB).  For every axion, only
one $U(1)$ can become massive; if there are more $U(1)$ factors arising from
branes than there are axions, the remainder must be non-anomalous.  Generally, the
non-anomalous $U(1)$'s will not simply be the diagonal subgroup living on a
brane stack.  Instead, they will arise as a linear combination of
various diagonal $U(1)$'s.
But nothing in our inflation scenario depended on the fact that each $U(1)$
lived on a brane.  We only utilized the fact that for every pair of gauge
groups there would generally be matter charged under both groups, and this will
be true for linear combinations of these groups as well.
So the picture above can go forward as before with each of the $U(1)$'s
having no mixed anomalies and arising from a linear combination of the
diagonal $U(1)$'s living on the various brane stacks.
As long as
there exists the
appropriate fields with charges to be the inflaton and waterfall (which
should arise just as easily for a $U(1)$ which arises from a linear combination),
the story goes through as before.

\subsection{Preheating and the Gravitino Problem}

As was shown
in \cite{Felder:2000hj,Felder:2001kt}, the
condensation of the waterfall field which ends hybrid inflation is typically
accompanied by tachyonic preheating (as opposed to the usual parametric resonance) in which
energy is dumped into the long-wavelength modes of the waterfall field itself.
Light scalar fields interacting with the waterfall field will be dragged along into the
process of preheating.  After a while, we expect these excitations to decay to a bath of fermions
and bosons in equilibrium.  Since this whole process is happening in a hidden sector, the reheating
could be happening in multiple stages and it could take a while (for sufficiently small coupling)
to reach the final relativistic bath of Standard Model particles.

This would in turn allow us to lower the temperature of reheating.  This is usually needed to avoid
the gravitino problem, in which
a low mass gravitino is overproduced at the end of inflation and
overcloses the universe. This is a serious problem to any theory of 
inflation with a high scale \cite{yanagida}. It
is naturally solved when the reheat temperature is roughly lower than $10^9 \sim 10^{10}$ GeV for
TeV scale supersymmetry breaking.
This can be accomplished very simply by having a long multi-stage reheating
process where the inflationary energy goes through multiple stages before settling
into its final Standard model bath of particles.  The details of the reheating process are model
dependent as one would need to know exactly how strong are the coupling between the inflaton sector
and the Standard Model branes.

Finally, it is possible that density perturbations and/or non-gaussianities can be generated
in the preheating era.  These were shown to be possible by using second order
analysis in \cite{Barnaby:2006cq}. Alternatively, it has been argued that the presence of
extra light fields on which the mass of the waterfall field
depends can also generate density perturbations at
the end of inflation \cite{Lyth:2006nx}.  All of this interesting issues are beyond the scope
of the current work and we assume that they change the inflationary prediction
only in a subdominant way.

\section{Hidden Sector Baryogenesis with Tachyonic Preheating}

This scenario for inflation in intersecting brane models can also
provide input for baryogenesis.  One of the most interesting models
of baryogenesis is electroweak baryogenesis \cite{EWBG}, in which the mixed
anomaly between $U(1)_B$ and the weak group $SU(2)_L$ creates a
non-trivial divergence in the baryon current:
\be
\partial_{\mu} J_B ^{\mu} \propto Tr \, [F_W \wedge F_W]\; .
\ee
Electroweak sphaleron processes can thus lead to microscopic
baryon generation.
If the electroweak phase transition is first order, then this
transition provides the departure from thermal equilibrium necessary
for the generation of a net baryon asymmetry.

One of the reasons for the great interest in this mechanism
is that it is determined
by physics at accessible (or soon to be accessible) energy scales and
thus can make definite predictions which can be tested.  In this regard
it is almost too successful; EWBG in the Standard Model is already ruled
out by LEP data, and in the MSSM it is tightly constrained into a very narrow
window.  These constraints largely arise from the necessity for the
EWPT to be first order: in the Standard Model one needs $m_{higgs} < 70 $ GeV,
and in the MSSM one needs $m_{higgs} < 120 $ GeV (the bounds from LEP establish
$m_{higgs} > 114 $ GeV).  In addition, the requirement that sufficient
CP violation be provided by the Higgs sector provides further constraints
(for an overview of all of these constraints, see \cite{Balazs:2005tu} and
references therein).

Thus, an interesting offshoot proposal of EWBG has been electroweak
tachyonic preheating \cite{EWTpreheat}.  In this model, the
electroweak phase transition
is a cold transition which occurs at the end of hybrid inflation.  The
waterfall field which ends inflation is the Higgs field, which simultaneously
breaks the electroweak symmetry.  Such a fast cold transition will necessarily
occur outside thermal equilibrium.
Furthermore, tachyonic preheating \cite{Felder:2000hj,Felder:2001kt,EWTpreheat} will robustly
dump energy into the long-wavelength modes of the fields coupled to the waterfall
field as it condenses.
These modes can thus excite the electroweak
sphalerons necessary for the generation of a baryon asymmetry.  This proposal
is a novel alternative which offers a chance to realize EWBG despite the
constraints on the EWPT.  However, it necessarily requires low-scale inflation
(in fact, electroweak scale), which is accompanied by its own set of constraints
in order to generate the correct scale of density perturbations.

On a parallel track, a recently proposed model of hidden sector
baryogenesis \cite{Dutta:2006pt}
realizes a mechanism reminiscent of EWBG in a context very natural to intersecting
brane models.  In HSB, mixed anomalies between a hidden sector gauge group $G$ and
$U(1)_B$ result in a non-trivial divergence for the baryon current
\be
\partial_{\mu} J_B ^{\mu} \propto Tr \, [F_G \wedge F_G]\; .
\ee
As a result, sphaleron/instanton processes in the hidden $G$ sector
can generate a baryon asymmetry in much the same way that electroweak sphalerons
generate it in EWBG.  These hidden sectors are
numerous and appear generically in intersecting brane models, and they generically
have mixed anomalies with both $U(1)_B$ and $U(1)_{B-L}$.  Since any
one of these groups can generate a baryon asymmetry if it has a strongly
first order phase
transition, and it only takes one for the mechanism to work, HSB provides a
robust mechanism for generating a baryon asymmetry in IBMs.  The generic appearance
of $[U(1)_{B-L}G^2]$ mixed anomalies implies that HSB generates a $B-L$ asymmetry
as well, which cannot be washed out by electroweak sphalerons.

We now see that HSB fits naturally into our $D$-term IBM inflation story
through tachyonic preheating.  The waterfall field $\phi_+$
is charged under $G_b$ and $G_c$, the gauge groups living on brane
stacks $b$ and $c$.  As it condenses, the group $G_b \times G_c$
will break to a subgroup $G$.  Via tachyonic preheating,
energy will be dumped into the long-wavelength modes of fields coupled
to $\phi_+$, and in particular into the gauge fields of $G_b \times
G_c$, which are simultaneously becoming massive.  As mentioned
earlier, brane stacks $b$ and $c$ will generically intersect with the
visible sector brane stack carrying $U(3)_{qcd}$.  The diagonal $U(1)$
subgroup of this will be $U(1)_B$, and thus we will generically have a
non-zero mixed anomaly of the form $[U(1)_B G_{b,c}^2]$ (in fact, there will
generically be a $[U(1)_{B-L} G_{b,c}^2]$ mixed anomaly as
well\footnote{Since the baryonic and leptonic stacks need not be parallel,
they generically have different topological intersection numbers with hidden
sector branes, thus leading to different coefficients for the $U(1)_B G^2$ and
$U(1)_L G^2$ mixed anomalies.  Note that this will not be the case in Pati-Salam
constructions, in which the baryonic and leptonic branes are parallel.}).
We may write this as
\be
\partial_{\mu} J_B ^{\mu} \propto C_b Tr[F_{G_b} \wedge F_{G_b}]
+C_c Tr[F_{G_c} \wedge F_{G_c}]\; .
\ee
Long wavelength modes of the
$G_b \times G_c$ gauge fields will excite sphalerons of the hidden sector, which will
in turn generate a baryon asymmetry via the anomaly, as in HSB (and analogous
to electroweak tachyonic preheating baryogenesis).  Moreover, this transition
will occur at the symmetry breaking scale of the $G_b \times G_c$ sector, which can be
relatively high.  So in this scenario, inflation does not need to be at a very
low scale, avoiding one of the constraints of electroweak tachyonic preheating.

A prototype for this might be an example where $G_b = U(1)$ and $G_c = U(2)$ or
$Sp(1)$, which
would be very similar to electroweak breaking.
In the original works on electroweak tachyonic preheating, numerical simulations
demonstrated that when the electroweak gauge theory
is broken by condensation of the waterfall
field, tachyonic preheating would indeed dump
energy into the gauge field, which in turn would
excite sphalerons and (accompanied by C and CP violation) generate a baryon
asymmetry.  Such a mechanism can certainly work in the HSB context
for the gauge theories arising in
intersecting brane models as well, provided that C and CP violation are present
(they should appear generically in the hidden sector gauge theory; whether
or not they are of the correct magnitude is a detailed model dependent
question which one must answer hidden sector by hidden sector).  But one still
must ensure that after tachyonic preheating, when the radiation has thermalized,
the sphalerons are shut off.  Otherwise, the sphalerons could wash out
the generated baryon asymmetry.  An estimate for the condition for this washout to
be suppressed is
\be
{T_{RH} \over \langle \phi_+ \rangle} < 1
\ee
where $T_{RH} \sim ({30 V_0\over \pi^2 g_*})^{1\over 4}$ is the (maximum) reheat temperature
and $g_*$ is the number of relativistic degrees of freedom into which energy
can be dumped.  For our case, we have $V_0 = g^2 \xi^2$.  Since we also have
$|\phi_+|^2 = \xi$ after condensation, we get
\be
{T_{RH} \over \langle \phi_+ \rangle} \sim {1.3 \over g_* ^{1\over 4}}
g^{1\over 2}
\ee
For $g_* \sim 10$ and $g \leq 0.9$, sphalerons will certainly be suppressed after tachyonic
preheating, and the generated asymmetry will not be washed out.

One might instead consider the case where $G_{b,c}$ are both abelian groups.  In this
case the hidden sector will
no longer have the complicated vacuum structure which allowed the sphaleron processes,
which in turn generated the baryon asymmetry via the anomaly.
An intriguing alternative would be the
non-equilibrium generation of charged particle/anti-particle pairs
and currents in the hidden sector.  For a $U(1)$ hidden sector gauge group, we have
\be
\partial_{\mu} J_B ^{\mu} \propto \int F_G \wedge F_G = \int E_G \cdot B_G
\ee
As energy from the waterfall field is dumped into the hidden sector, one expects
the generation of charged particle/anti-particle pairs which would in turn
lead to non-trivial hidden sector electric and magnetic fields.  The lack of
thermal equilibrium, and violation of B, C and CP imply that all Sakharov
conditions are satisfied, and thus that one can generate
a baryon asymmetry through the anomaly.  But it is a non-trivial task to understand
the precise mechanism by which departure from
equilibrium, combined with C and CP violation, can generate the
required non-zero $\int E \cdot B$.  Furthermore, it would be a
technical exercise to verify that, as the gauge field becomes massive during
the waterfall stage, the electric and magnetic fields are suppressed
sufficiently quickly to shut-down the baryon-violating processes before
thermalization, in order to avoid washout of the asymmetry.  These interesting
questions are beyond the scope of this work, though we hope to revisit them.

\section{The Generation of  Neutrino Masses}
The condensation of the waterfall field breaks a gauge symmetry and
the scale of this symmetry breaking can be around
$10^{14}-10^{16}$GeV. This gauge symmetry can thus be used to protect the
Majorana mass of right handed neutrino and this mass scale can be
used to generate  light neutrino masses via the seesaw
mechanism~\cite{seesaw}. The light neutrino mass can be written as
$m^{\rm light}_{\nu}=-M^D_{\nu}{1\over M_{Maj}}{M^D_{\nu}}^T$, where
$M^D_{\nu}$ is the Dirac neutrino mass which arises from the
$\lambda_{\nu}L\nu^c H$ coupling and the magnitude of this mass is
typically of the order of the weak scale. The Majorana mass arises
from the $f\nu^c\nu^c \phi$ interaction and the magnitude of this
mass is of the order of the symmtery breaking scale where the MSSM
is generated. After including the magnitudes of these mass scales in
the seesaw formula, one can find that the heaviest light neutrino
mass can be 0.1 eV or less. The precise values and the flavor
structures of the Dirac and Majorana couplings will generate the
neutrino mixing angles and masses to satisfy the experimental
results.  Here, we are interested to see how the light neutrino
masses are connected to our inflation mechanism.

We can consider this scenario in an effective field theory approach.
Here, the field $\phi$ which appears in the Majorana interaction is the waterfall
field. The full gauge symmetry  is protected until the waterfall field
condenses and this symmetry  can be, e.g., $SU(3)_c\times
SU(2)_L\times U(1)_{T_{3R}}\times U(1)_{B-L}$, with
both $\phi$ and $\nu^c$  are charged under
$T_{3R}$ and ${B-L}$.
$U(1)_{T_{3R}}\times U(1)_{B-L}$ is then broken down to $U(1)_Y$ due to
the condensation of $\phi$.
$T_{3R}$ charges are assigned to the fields in
such a way so that fields have the correct hypercharge. The Majorana
mass term develops once the waterfall field  develops a VEV. In this
process both $B-L$ and $T_{3R}$ symmetries are spontaneously broken
and the baryon asymmetry that is produced due to the $[U(1)_{B-L}G^2]$
anomaly is not washed out due to electroweak sphaleron process.

\begin{figure}[ht]
\centering
\includegraphics[width=8cm]{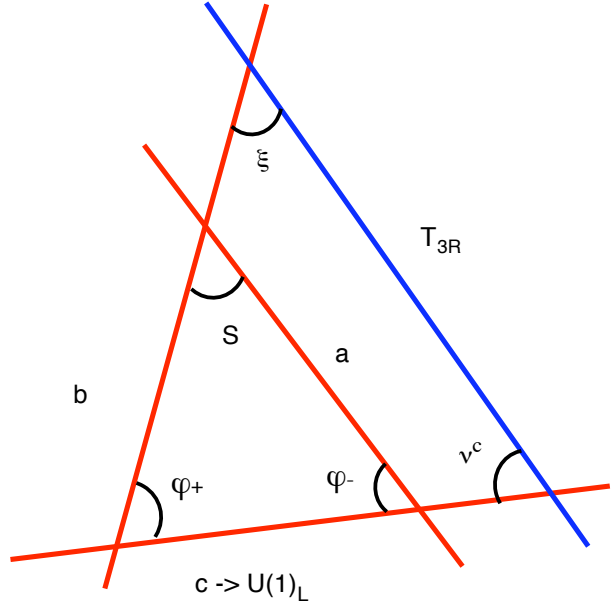}
\caption{One can simply generate the neutrino masses at the end of
inflation by making stack $c$ the leptonic brane with
gauge symmetry $U(1)_L$. We then need both stacks $b$ and $c$ to
intersect the $T_{3R}$ brane in order to generate the correct
Yukawa coupling.  Here in addition to the previous inflaton sector fields,
we have the right-handed
neutrino $\nu^c$ and an exotic $\xi$.}\label{basicplusneutrino}
\end{figure}

We can also consider this scenario in an IBM implementation of $D$-term
inflation which we have discussed here.
For example, we consider the case where the $c$ brane stack
which we discussed earlier is in fact the leptonic stack, on which
$U(1)_L$ lives.  Generically we expect the $T_{3R}$ stack to have
non-vanishing topological intersection with the the hidden sector
stack $b$ (on which $\phi = \phi_+$ also ends, see Fig. \ref{basicplusneutrino}).
Assuming that the
right-handed neutrinos arise from multiplets living on the intersection
of the leptonic and $T_{3R}$ brane stacks, we now find that allowed
renormalizable Yukawa couplings are
\be
W_{yuk.} = \lambda S \phi \phi_- + f \phi \xi \nu^c
\ee
If we denote by $U(1)_b$ the diagonal subgroup of $G_b$, then we
can define the hypercharge group $U(1)_Y = U(1)_{B-L} + U(1)_{T_{3R}}
+U(1)_b +...$, under which SM particles have correct hypercharge and
$\phi$ and $\xi$ have vanishing hypercharge.
The extra Yukawa coupling will not affect the inflation potential significantly,
as the scalars in the $\xi$ and $\nu^c$ multiplets have vanishing
expectation value during inflation.
$\xi$ is a Majorana field whose mass is
protected by the $T_{3R}$ symmetry which can also break around the
scale where $B-L$ breaks down.
The Majorana mass of $\nu^c$ arises from a tree-level interaction with
the fermionic component of $\xi$ appearing as the intermediate line.  We
thus find $m_{\nu^c } \sim f^2 \langle \phi \rangle^2/M_{\xi}$,
where $M_{\xi}$ is $\sim$ GUT scale and is, therefore, of the right
magnitude to generate the  light neutrino mass.

It is interesting to note that lepton asymmetry in this model can be
generated by CP-violating out-of-equilibrium decays of right handed
neutrinos~\cite{lepto}. This asymmetry is obtained via the
tree and loop contributions in the decay of the lightest
right-handed Majorana neutrino into leptons and scalar Higgs. The
electroweak sphaleron process then converts the lepton asymmetry to a
baryon asymmetry.

Although hidden sector baryogenesis and the generation
of mass for the neutrinos can both fit nicely and simultaneously into
this picture of $D$-term inflation in IBMs, it is not necessary for
both mechanisms to appear simultaneously.  For example, we can consider
IBMs where the baryon asymmetry is generated at the end of inflation by
HSB associated with tachyonic preheating, while the neutrino masses are
generated by unconnected dynamics.  Similarly, we can consider models where
the condensation of the waterfall field generates the neutrino masses, but
baryogenesis is dominated by contributions other than HSB (including perhaps
leptogenesis of the form discussed above).

\section{Conclusion}

We have developed a scenario of $D$-term inflation which appears to
arise quite straightforwardly in the context of intersecting brane
models.  In this case, the inflaton $S$ lives in a chiral multiplet
arising from strings stretching between two hidden sector branes at
their point of intersection.  The waterfall field has a similar
origin.  A noteworthy feature of this type of inflation is that it
can quite naturally generate a baryon asymmetry in the visible
sector at the end of inflation via the recently proposed hidden
sector baryogenesis mechanism \cite{Dutta:2006pt}.  One can also
easily generalize this model to additionally generate Majorana
masses for neutrinos.  One of the puzzles of cosmological evolution
has been the origin of the various required scales.  One would need
to understand where the scale of the Majorana masses arises from,
and depending on the mechanism of baryogenesis, also the origin of
the scale at which one departs from equilibrium.  The scenario we
discuss here ties both of these scales directly to the scale of
inflation; to our knowledge, it is the only scenario to achieve
this. Further, both high and low scale supersymmetry breakings are
compatible with  baryogenesis which can happen in the hidden sector or
via leptogenesis.

Although a key motivation for this scenario comes from the general
features of intersecting brane models, this scenario can be thought of
purely in terms of effective field theory.  As such, it may have wider
application.  Of course, we think of this as a scenario instead of a
true ``model" since we have not illustrated with a specific construction.
Thus far, it has proven relatively difficult to generate specific
examples of intersecting brane models in compactifications in which all
closed string moduli are fixed, although some work on this subject has
been done \cite{IBMFV}.  There are many difficulties in combining an
intersecting brane model with the features necessary for fixing moduli,
such as fluxes and non-perturbative corrections, but it seems to us that these
problems are technical in nature; there is no clear reason why shouldn't
be able to stabilize moduli in some compactification with an IBM.

More specifically, there are two main working assumptions in this paper that remain to
be worked out in detail in order to achieve a consistent string theory embedding of this
scenario.  One assumption is  that there exists a consistent
moduli stabilization scheme where the $D$-term energy dominates over the $F$-term.  
Secondly, we have not included any quantum corrections to the  FI term which may depend on the inflaton field. 
One might expect that  the quantum
corrections are suppressed since the inflaton field is a bifundamental living at an intersection separated from the brane that generates the dominant FI term during inflation.
We leave an investigation of these issues for future work~\cite{wp}.

To the
extent that one thinks of this as inflation in an intersecting brane
model (as opposed to a more generic effective field theoretic setting),
it obviously relies on the existence of many such moduli-stabilized
IBMs.  It would of course be very interesting to understand specific
examples of such moduli-stabilized IBMs, and to see if this type of
inflation scenario can be realized.  Along these lines, another very
interesting problem is
to better understand which branes and features are generic in the
class of moduli-stabilized intersecting brane
models, beyond the more well-understood set of toroidal orientifold
constructions.

\section*{Acknowledgements}
We gratefully acknowledge P. Batra, M. Dine, S. Kachru, J. Kratochvil, L. McAllister,  R. McNees, D. Morrissey,   S. Sarangi,
S. Shandera, G. Shiu, H. Tye, G. Villadoro and J. Wells for useful discussions.
JK is grateful to the KITP for
hosting the ``String Phenomenology" workshop, where some of this
work was done.  LL is grateful to the ISCAP institute where part of this work was done.
This work is supported in part by
NSF grant PHY-0314712 and PHY-0555575.

\end{document}